\documentclass[
 reprint,
 amsmath,amssymb,
 aps,
pra,
]{revtex4-2}

\usepackage{graphicx}
\usepackage{dcolumn}
\usepackage{bm}

\begin{document}

\title{Non-ideal subthreshold swing in aligned carbon nanotube transistors due to variable occupancy discrete charge traps}

\author{Saurabh S. Sawant}
\affiliation{
 Center for Computational Sciences and Engineering, Lawrence Berkeley National Laboratory, Berkeley, California, USA}

\author{Teo Lara}
\affiliation{
 Department of Physics, Massachusetts Institute of Technology, Cambridge, Massachusetts, USA}
 \affiliation{
 Center for Computational Sciences and Engineering, Lawrence Berkeley National Laboratory, Berkeley, California, USA}

\author{Fran\c{c}ois L\'eonard}
\email[Contact author: ]{fleonar@sandia.gov}
\affiliation{
 Sandia National Laboratories, Livermore, California, USA}
 
\author{Zhi (Jackie) Yao}
\affiliation{
 Center for Computational Sciences and Engineering, Lawrence Berkeley National Laboratory, Berkeley,California, USA}

\author{Andrew Nonaka}
\email[Contact author: ]{ajnonaka@lbl.gov}
\affiliation{
 Center for Computational Sciences and Engineering, Lawrence Berkeley National Laboratory, Berkeley,California, USA}

\date{\today}

\begin{abstract}
Carbon nanotube transistors have been experimentally demonstrated to reach performance comparable and even surpassing that of silicon transistors. Further improvement requires addressing non-idealities arising from device fabrication that impact performance and reproducibility. One performance metric that determines energy efficiency is the subthreshold swing which is often observed to be 3-4 times larger than the ideal thermal limit. In this work, we present simulations indicating that a discrete number of variable occupancy hole trapping sites can explain the large subthreshold swing. Our simulations indicate that while three-dimensional trap distributions influence the subthreshold swing, only the traps in close proximity to the nanotubes have a significant impact. The results suggest that a density of trapping sites on the order of 0.5/nm$^2$ near the nanotubes is sufficient to significantly increase the subthreshold swing, requiring the removal or passivation of only a few sites per carbon nanotube.
\end{abstract}

\maketitle

\section{Introduction}\label{sec:introduction}

Carbon nanotube transistors (CNTs) show promise for next-generation computing devices \cite{Shulaker2013, Han2017} as well as for applications in optoelectronics \cite{Bergemann2018, Bergemann2020}.
Progress in purification and alignment of carbon nanotubes as well as optimization of contacts and dielectrics has led to significant performance improvements \cite{Brady2016,Lin2023,Safron2024}.
However, there remain issues with device variability and performance behavior that need to be addressed to reach ultimate performance.
While several mechanisms have been suggested for device imperfection (e.g. nanotube diameter distributions, nanotube misalignment), these mechanisms are unable to explain the unusually large subthreshold swing ($SS$) observed in some devices \cite{Lin2023, Lin2023b}. $SS$ is the measure of how many millivolts of gate voltage are required to increase the transistor current by one decade in its subthreshold region, and it is critical for achieving lower power consumption by allowing the device to switch with minimal voltage swing.
Indeed, while the thermal limit \cite{sze_physics_2007} suggests that $SS = 60$ mV/decade is possible at room temperature, several experiments report values in the range of $200$ mV/decade.
In addition, measured values of $SS$ can vary within the same device preparation methodology \cite{Cao2015}, and between different fabrication approaches \cite{dna-assembly}.
Reducing $SS$ is important to lower power consumption in digital logic~\cite{Truesdell2020} and to increase the responsivity of sensors based on field-effect transistors (FETs)~\cite{Wang2023}.

\begin{figure*}
\centering
\includegraphics[width=0.75\textwidth]{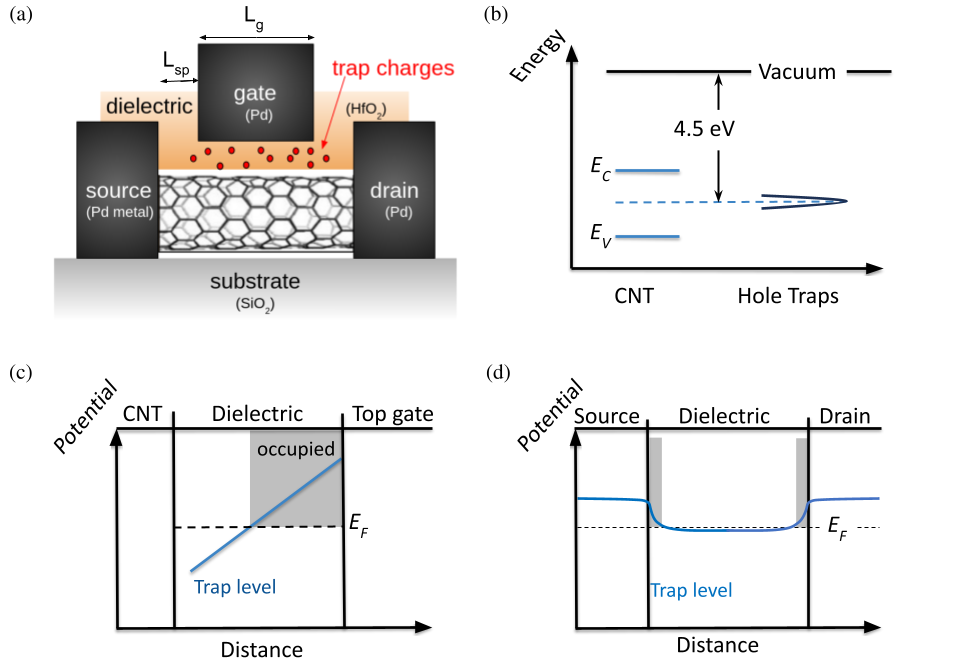}
\caption{Model for the charge traps in the gate oxide for a top gate CNTFET configuration. (a) Geometry of the top gate device under consideration. (b) Energy level of the trap with respect to the vacuum level and the CNT conduction and valence band edges. (c) Illustration of the variation in the trap energy between the CNT and top gate when a gate voltage is applied. (d) Illustration of the variation in the trap energy in the dielectric in a direction along the CNT.}
\label{f:Fig1}
\end{figure*}

Trapped charges are a possible culprit for the increased $SS$ since they screen the gate electric field and reduce its effectiveness.
The role of trap charges in CNTs has been addressed in two ways.
In a continuum picture a uniform sheet of charge is predicted to change the device capacitance and give
\begin{equation}
    SS = SS_0 \left(1 + \frac{C_D + e^2 N_{\rm it}}{C_{\rm ox}} \right),
    \label{eq:SS_1}
\end{equation}
where $SS_0 = \ln(10) k_b T / e$ is the thermal limit, $k_b$ is Boltzmann's constant, $T$ is the temperature, $e$ is the elementary charge, $C_D$ is the depletion capacitance per area, $C_{\rm ox}$ is the gate capacitance per area, and $N_{\rm it}$ is the trap interface state density (states per area per eV).
For a long-channel device with a thin gate, $C_D$ can be neglected, and we can use the standard expression $C_{\rm ox}=\epsilon / t_{\rm ox}$ to obtain
\begin{equation}
    SS = SS_0 \left(1 + \frac{e^2N_{\rm it}t_{\rm ox}}{\epsilon} \right).
    \label{eq:SS_2}
\end{equation}
Here $\epsilon$ is the oxide dielectric constant and $t_{\rm ox}$ its thickness.
Application of this simplified expression (or a combination of techniques to measure $C_D$ and $C_{\rm ox}$) allows the extraction of $N_{\rm it}$ for experimental devices.
Note that this quantity is both continuous in the density of trapped sites as well as in the density of electronic states.
Using this continuum approach, values of the state density around $10^{12}$ cm$^{-2}$eV$^{-1}$ have been obtained from experimental data \cite{liu2024interface} for CNTFET top gate geometries.
However, determining the actual number of traps is difficult because of the inherent assumption of a continuous density of trap states.
For example, the total number of charges would be $N_{\rm tot}=N_{\rm it} L_{\rm g} W \Delta E$ where $L_{\rm g}$ is the gate length, $W$ is the channel width, and $\Delta E$ is the difference between the Fermi level and the bottom of the interface state distribution.
For a device with $L_{\rm g} = W = 100$~nm and the above value of $N_{\rm it}$ we have $N_{\rm tot}=(100~ eV^{-1})\Delta E$.
Taking $\Delta E=1$~eV, we obtain $N_{\rm tot}=100$, a discrete number of charges even for this moderate device size. Further reduction to a 10 nm gate length would give even fewer trap sites and the assumption of a continuous distribution of states must be revisited.
In addition, the unknown value of $\Delta E$ leads to uncertainty in the total number of charges.
Furthermore, the above expression also assumes that traps have the same impact on the channel regardless of their position.
In the case of CNT arrays where the pitch is typically a few times the CNT diameter, the proximity of traps to the CNT varies and determines their impact on performance.

An alternative approach is to consider the impact of discrete charges on CNT device performance. 
Initial work focused on traps with fixed charge (i.e. charge independent of gate voltage). 
A priori, fixed charges can explain shifts in threshold voltage but cannot account for non-ideal subthreshold swings based solely on electrostatic arguments.
Indeed, the above continuum model applied to a sheet of {\it positive} fixed charges gives an electrostatic potential on the CNT:
\begin{equation}
    V_{\rm CNT} = V_{\rm G} + \frac{e N_{\rm it} \Delta E t_{\rm ox}}{\epsilon},
    \label{eq:V_CNT}
\end{equation}
which can be viewed as a shift in the gate voltage but does not change the proportionality constant between $V_{\rm CNT}$ and $V_{\rm G}$ that would be necessary to change the $SS$.
However, simulations of gate-all-around CNT transistors with a single fixed charge showed that for SiO$_2$ as the dielectric the $SS$ was increased by 30\% from the thermal limit for a charge 4 \AA ~from the CNT surface~\cite{wang2007random}.
This effect is a purely quantum effect arising from a tunneling potential created by the charge and an energy-dependent electron transmission.
Subsequent modeling work considered this effect in the case of a single CNT on planar SiO$_2$ with fixed charges randomly deposited on the SiO$_2$ surface~\cite{Cao2015} utilizing a simplified model of electronic transport in the CNT to assess the impact on transistor performance. 
While a strong effect was predicted (perhaps due to the weak dielectric environment) the accuracy of the simpler model, and its applicability to emerging CNT device geometries need to be assessed due to several developments in CNT device fabrication. First, the needed ON-state current density for electronics applications requires dense arrays of aligned CNTs instead  of individual CNTs. 
Second, the demonstration of top-gate geometries opens the possibility of three-dimensional distributions of trapped charges.
Third, approaches for separating and assembling CNTs with surfactants can introduce residual sites for charge trapping.

In this work, we present simulations for emerging CNT device geometries using a model for trap charges that bridges the discrete and continuum approaches by allowing the occupancy of discrete energy-localized charge traps to depend on the local electrostatic potential.
Comparison with experimental data suggests that the number of relevant occupied charge traps increases linearly with gate voltage reaching around 0.5/nm$^2$ in the ON-state. The results imply that a scaled device with 10 nm channels and 2 nm separation between CNTs would require the removal or passivation of only a few traps per CNT.

\section{Methods}\label{sec:methods}

We focus on comparison with recent experimental results for 3D top gate geometries~\cite{Lin2023} with densely aligned CNTs. 
Figure~\ref{f:Fig1}a shows a schematic of the device.
The device consists of aligned CNTs of 1.57~nm diameter and 2.556~nm pitch sitting on SiO$_2$ ($\epsilon=3.9$) and covered with 5~nm HfO$_2$ ($\epsilon=25$). 
A top gate of length $L_{\rm g}=85$~nm covers the central part of the channel with two spacer regions of length $L_{\rm sp}=5$~nm on either side of the gate, giving a total channel length $L_{\rm ch}=L_{\rm g} + 2 L_{\rm sp}=95$~nm.

\begin{figure}
    \centering
       \includegraphics[width=0.4\textwidth]{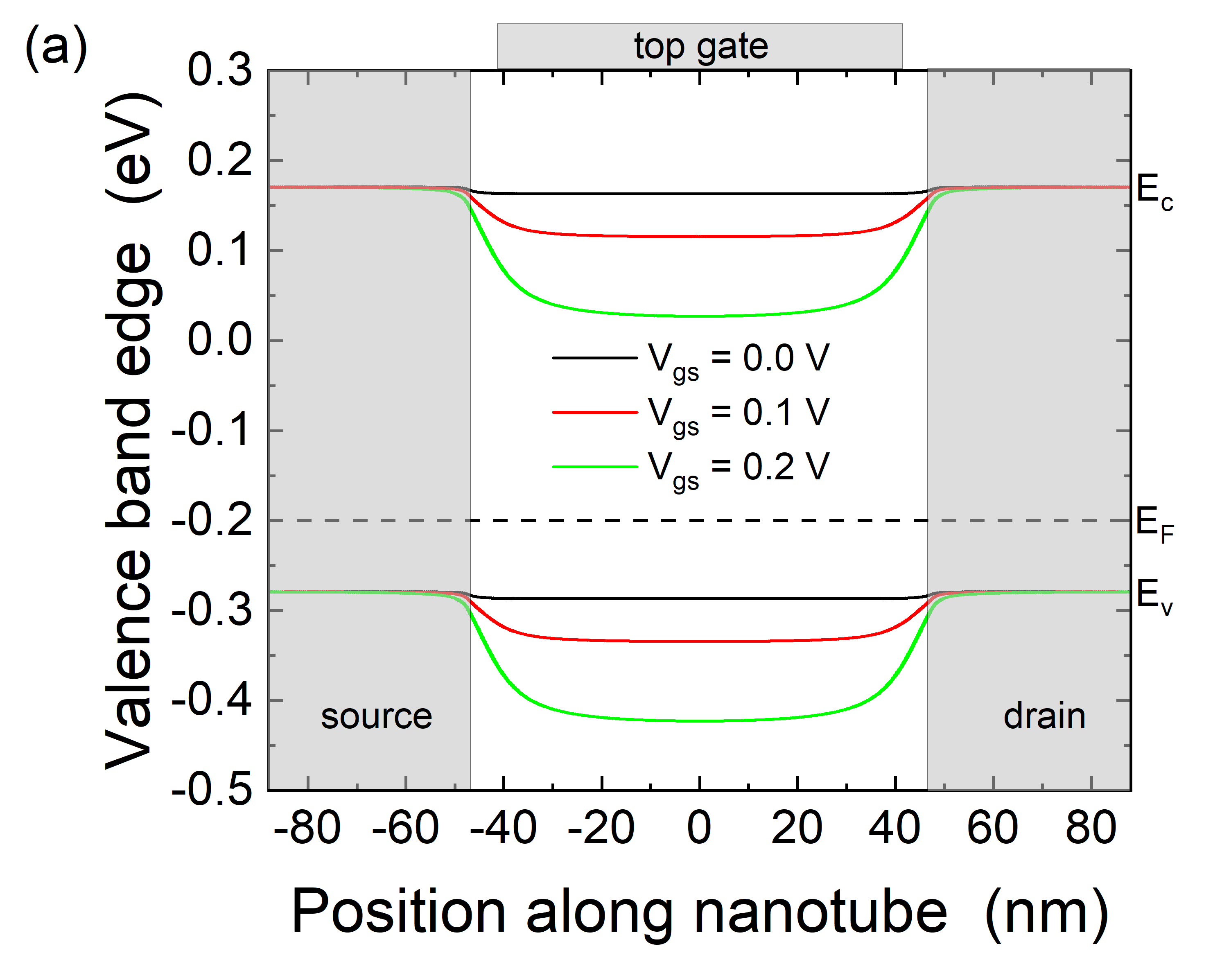}
       \includegraphics[width=0.4\textwidth]{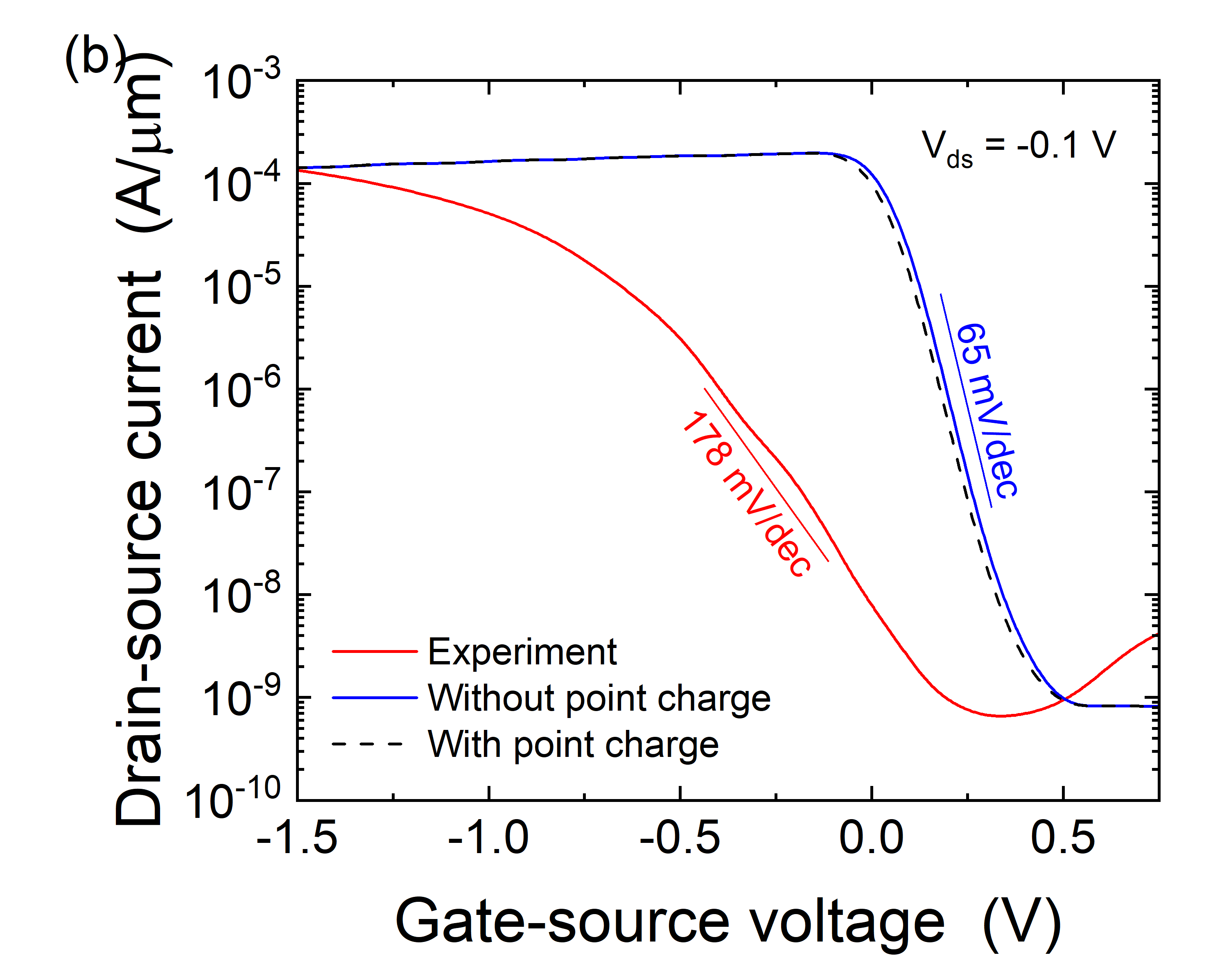}
       \includegraphics[width=0.4\textwidth]{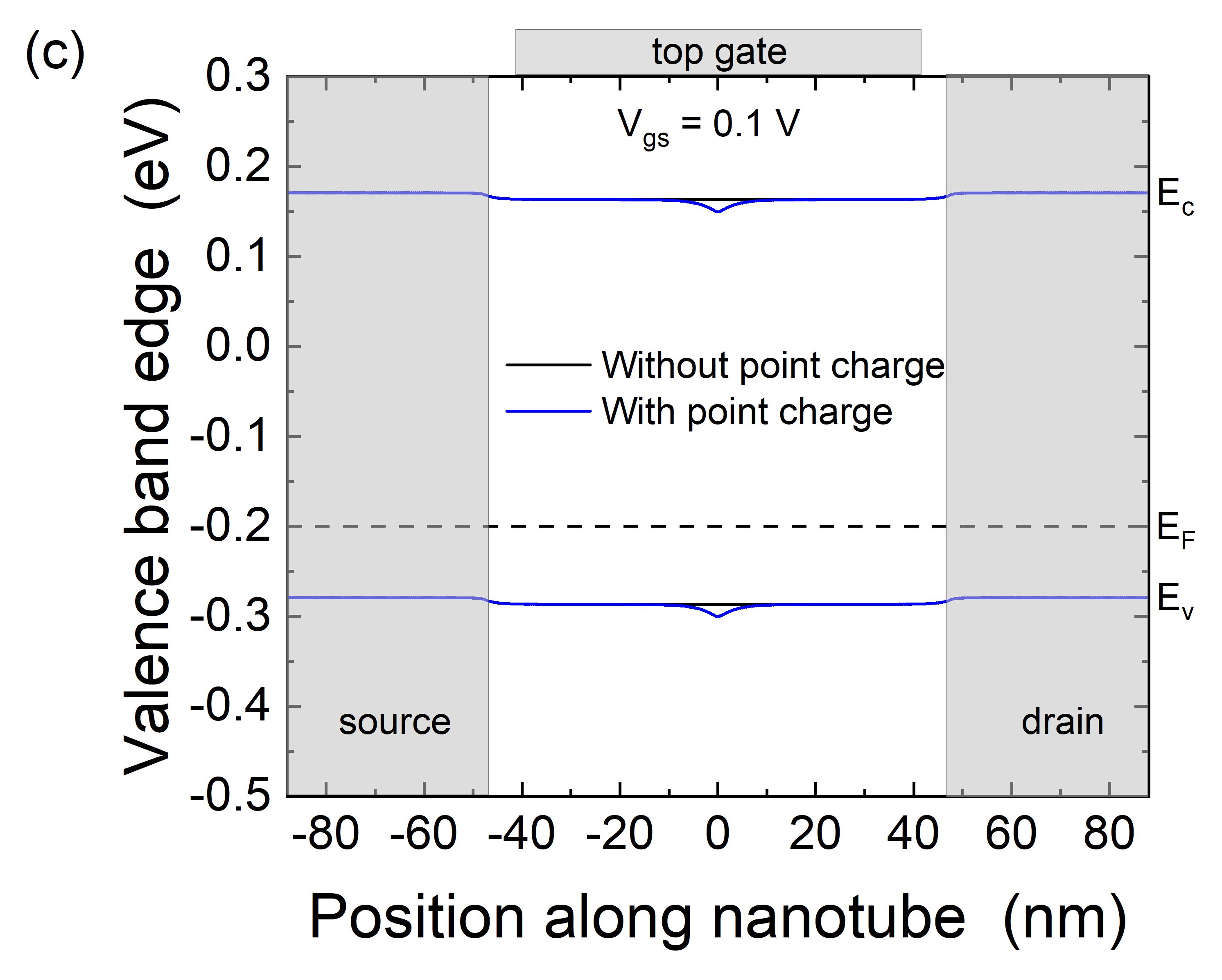}
       \caption
       {Comparison of the impact of a single charge versus no charge. (a) Band-bending across the CNT at three gate-source voltages. \textbf{b)} Transconductance along with experimental data~\cite{Lin2023}, showing subthreshold swing for each curve. (c) Band-Bending in the presence of a single charge compared with the case of no charge.}
    \label{f:Fig2}
\end{figure}

We model the CNT electronic properties using a tight-binding model with a nearest-neighbor overlap integral of 2.5 eV. 
We focus on (20,0) CNTs with a diameter of 1.57 nm and a bandgap of 0.45 eV.
The top of the SiO$_2$ and the bottom of the HfO$_2$ are planar and separated by 2.165 nm
of vacuum.
The aligned CNTs are centrally inserted in this vacuum region and spaced 0.3 nm away from both the SiO$_2$ and the HfO$_2$ interfaces.
In the contact regions, the CNTs are enclosed above and below by planar metal blocks of length 40~nm.
The CNT midgap 4.5 eV below the vacuum is chosen as the reference level and the contact Fermi level is chosen to be $E_F=-0.2$ eV in order to match the ON-state conductance and minimum current of the experimental devices, as discussed below. This choice is representative of metal contacts with Pd which have been shown to give near-ohmic contacts.

The three-dimensional device geometries with dense arrays of CNTs and randomly distributed discrete charges bring challenges for CNT device modeling.
To address these challenges, we utilize the ELEQTRONeX modeling framework \cite{sawant}, which is a recently developed massively-parallel GPU-accelerated non-equilibrium Green's function (NEGF) simulation approach.
Simulations with this code have shown that CNT misalignment or variations in array density are not the cause of the large non-ideal $SS$ observed experimentally~\cite{sawant}. 
Specifically, we previously demonstrated these effects are responsible for at most 12\% degradation in SS.
In this work, we incorporate Lagrangian particles to model discrete trapped charges.
These particles interact with the underlying structured mesh that contains a continuum representation of charge and electric potential by depositing charge and interpolating electric potential to and from the mesh.
The operations utilize the cloud-in-cell algorithm~\citep{birdsall1969_cic} as described in the supplementary note 1 of Ref.~\cite{sawant}.

The model for the charge traps is illustrated in Fig.~\ref{f:Fig1}. 
The traps are assumed to have the same narrow energy level located in the bandgap of the CNT.
When voltages are applied to the source, drain, and gate, the trap energy is shifted by the electrostatic potential at the trap location, leading to trap occupancies that vary in the vertical (Fig.~\ref{f:Fig1}b) and horizontal directions (Fig.~\ref{f:Fig1}c) since the hole traps are occupied (empty) if the trap energy is above (below) the Fermi level $E_f$.
The charge on each hole trap is then obtained from 
\begin{equation}
    Q_i = e \sigma\left(\frac{eV_o - eV_i}{eV_t}\right),
    \label{eq:Q_i}
\end{equation}
where $\sigma$ is the sigmoid function, $V_i = V(\textbf{r}_i)$ is the electrostatic potential for the trap located at position $\textbf{r}_i$, $eV_0$ is the trap energy level, and $V_t$ is a parameter to smoothen the transition between the empty and filled trap, which originates from broadening of the trap energy level.
In our simulations, we found that $V_0 = 0.5$ V and $V_t = 20$ mV gave the best results in terms of convergence and comparison with experiment. Since most CNTFETs turn on as a negative gate voltage is applied the sign of the trapped charge is chosen to be positive since it reduces the effective negative potential on the CNT compared to the applied gate voltage as shown in Eq.~(\ref{eq:V_CNT}).

For the system without charge traps, we employ a self-consistent Broyden scheme to converge the charge and the potential on the CNT atoms. In the presence of charge traps, we add an additional self-consistent scheme in parallel for the charge on the traps. For this, we use a simple mixing algorithm to update the potential at trap locations, i.e. at iteration $j$
\begin{equation}
    V_{i}^{j} = \eta V_{i}^{j-1} + (1-\eta) V_{i}^{\rm temp},
    \label{eq:SimpleMixing}
\end{equation}
where $V_{i}^{j}$ is the local potential at a given iteration, $V_{i}^{\rm temp}$ is the potential calculated with charge $Q_{i}^{j-1}$  and $\eta$ is a mixing factor, typically set to be less than 0.1.
At a given gate-source and gate-drain biases, we continue performing Broyden's iterations (see supplementary note 4 of Ref.~\cite{sawant}), however, the charge on the nanotube does not converge until the potential at the trap locations is also converged. In other words, the presence of traps slows down convergence; in the results below, convergence without traps takes on average 29 Broyden iterations, while the cases with 3D and 2D trap distributions take 48 and 59 Broyden iterations, respectively. Because CNT misalignment is not a major factor, we accelerate the simulations by focusing on parallel CNTs and utilize periodic boundary conditions with one CNT per cell.

\section{Results}\label{sec:results}
We first describe the transistor action in the CNT array without trap charges. Figure~\ref{f:Fig2}a shows the calculated self-consistent band-bending across one of the CNTs within the array in the absence of charge traps.
Deep in the contact regions, the Fermi level lies slightly above the valence band edge, giving a near-ohmic contact.
Within the channel, a small positive gate voltage of 0.1V gives nearly flat bands and a high conductance, whereas more positive gate voltages introduce a potential barrier that impedes carrier flow and reduces conductance.
This transistor action is shown in the $I-V_{gs}$ curve of Fig.~\ref{f:Fig1}b.
In this idealized scenario, the subthreshold swing approaches the thermal limit of 60 mV/decade. 
Figure~\ref{f:Fig1}b also includes experimental data from Ref.~\cite{Lin2023} that is markedly different from the ideal simulations, exhibiting a $SS$ of approximately 178 mV/dec.

\begin{figure}
    \centering
       \includegraphics[width=0.45\textwidth]{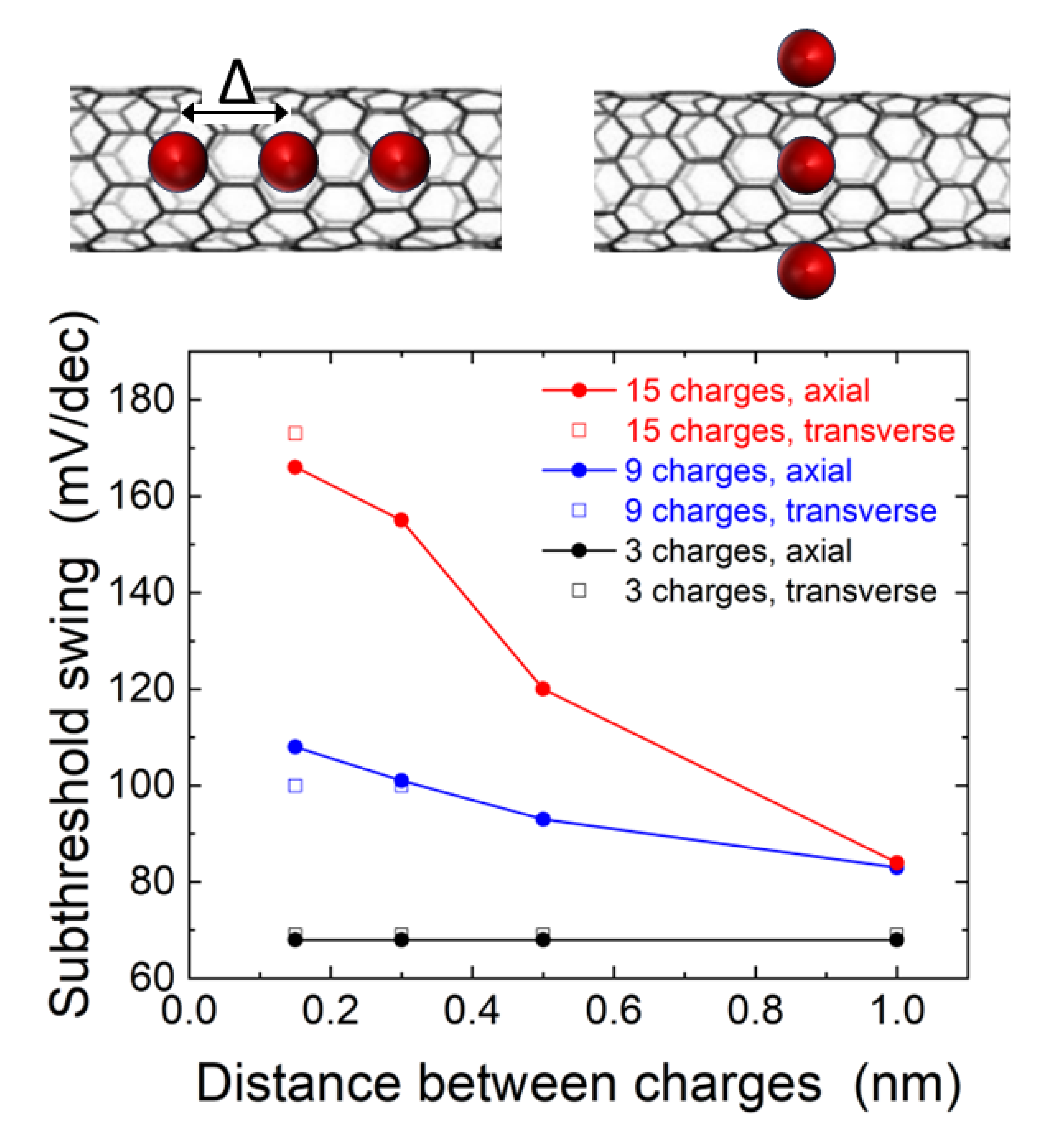}
       \caption
       {Impact of clusters of discrete charges on the subthreshold swing. The top illustrations show the axial and lateral configurations considered, with the spacing $\Delta$ between charges. The main panel shows the subthreshold swing as a function of $\Delta$ for clusters of 3, 9, and 15 charges.}
    \label{f:Fig3}
\end{figure}

To assess the importance of charge traps for this geometry, we first consider the case of a single fixed charge. 
Figure~\ref{f:Fig2}c shows the band-bending calculated in the presence of a single fixed charge located at a distance 0.4065~nm above the CNT in the center of the channel, which includes a 0.3 nm vacuum gap between the CNT and the HfO$_2$ interface. This distance was chosen to be as close to the CNT as possible while avoiding being too close to the boundary condition at the HfO$_2$ interface, and being commensurate with the simulation grid.

\begin{figure*}
    \centering
       \includegraphics[width=0.75\textwidth]{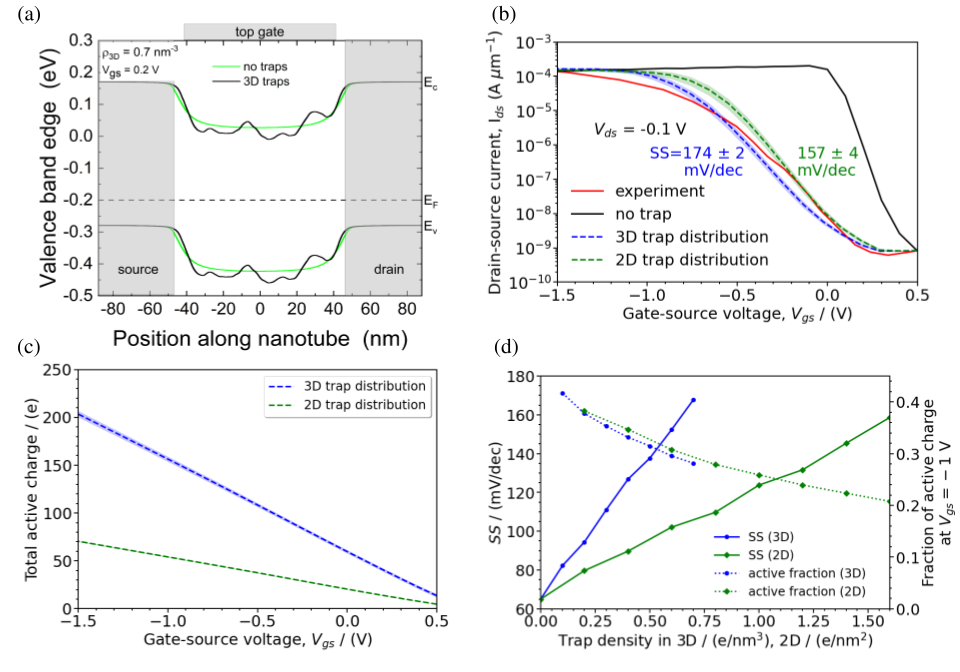}
       \caption
       {Effect of variable-occupancy 3D and 2D discrete random trap distributions. (a) Band-bending across the CNT for a 3D distribution. (b) Transistor transfer characteristics for 3D and 2D distributions compared with experimental data from Ref.~\cite{Lin2023} The thicker coloring around the dashed lines represents variations from 16 independent runs with different random configurations. (c) Number of occupied traps as a function of gate voltage for 3D and 2D distributions. (d) Effect of trap density on subthreshold swing, and fraction of active charge at $V_{gs}=-1$ V.}
    \label{f:Fig5}
\end{figure*}
We see a minute 0.01~eV dip in the band-bending at the center of the channel at $V_{gs}=0.1$~V, which does not change the $SS$ as seen from Fig.~\ref{f:Fig2}b. 
This differs from the results in Ref.~\cite{Cao2015} for various reasons.
First, the gate dielectric constant is much larger (25 vs 3.9) which screens the charge. Second, the use of the full self-consistent non-equilibrium quantum transport calculations captures more quantitatively the impact of the charge on the current.

Next we analyze the impact of multiple charges placed in a line on the $I-V_{gs}$ characteristics, considering two orientations--in the direction of the carbon nanotube axis (axial) and in a width-wise direction at the center of the channel (lateral), see illustrations in Figs.~\ref{f:Fig3}.
In both configurations, charges are located $0.4065$ nm above the nanotube, as in the single charge study.
We considered clusters of 3, 9, and 15 charges, with up to four different spacings between charges, $\Delta=$1, 0.5, 0.3, 0.15~nm. 
For the lateral orientation, only those spacings are considered for which the traps remain within the pitch of $2.556$~nm between CNTs in the array.

For three traps, Fig.~\ref{f:Fig3} reveals that the $SS$ changes very little, even when the traps are very close together. For nine charges the effect is greater, causing a more prominent increase of the $SS$ for closely spaced charges;
the axial orientation shows some degradation in $SS$ from 82 mV/dec to 113 mV/dec as the trap spacing decreases.
For the lateral orientation we consider the cases $\Delta=0.3$ nm and 0.15 nm, both exhibiting $SS$ = 100 mV/dec, similar to the axial orientation. Finally, for 15 traps we see substantial degradation in the $SS$ for $\Delta <=0.5$, with the $SS$ as high as 172 mV/dec for $\Delta=0.15$~nm with a similar result for the lateral orientation.

The conclusion from these simulations is that a single trap, or a cluster of a few traps, is not able to appreciably change the subthreshold swing. To see an appreciable impact close to the experimental $SS$ value, the cluster must consist of 15 or more traps that are very closely spaced (1.5 \AA).
For this mechanism to be played out in practice, we would need every CNT in the array to be impacted, i.e., for such a cluster to be present on each CNT in the array across the entire width of the channel. Furthermore, an ideal match occurs only at $\Delta=0.15$ nm, i.e. when the spacing between traps is as small as the bond length, in which case the material would be severely degraded and have poor dielectric properties.

As an alternative, we consider a situation where traps are more sparsely and randomly distributed in the gate oxide and that a different mechanism may be responsible for giving a gate-dependent potential.
In the continuum model, this comes from the constant density of trap states that get filled as the gate voltage changes.
As discussed in the context of Eq.~(\ref{eq:Q_i}) we capture this type of effect with discrete charges by allowing the occupancy of each charge to be filled according to the electrostatic potential at each charge. 
Here, we explore two ways of randomly distributing traps: traps with a 3D distribution, where they may be located within the entire gate oxide region in the portion lying below the gate, and traps with a 2D distribution, where they lie on a planar sheet located $0.4065$~nm above the nanotube. 
Next, we show that as the random trap density is increased, we see a similar degradation of the $SS$ as observed experimentally.
\begin{figure*}
    \centering
       \includegraphics[width=0.75\textwidth]{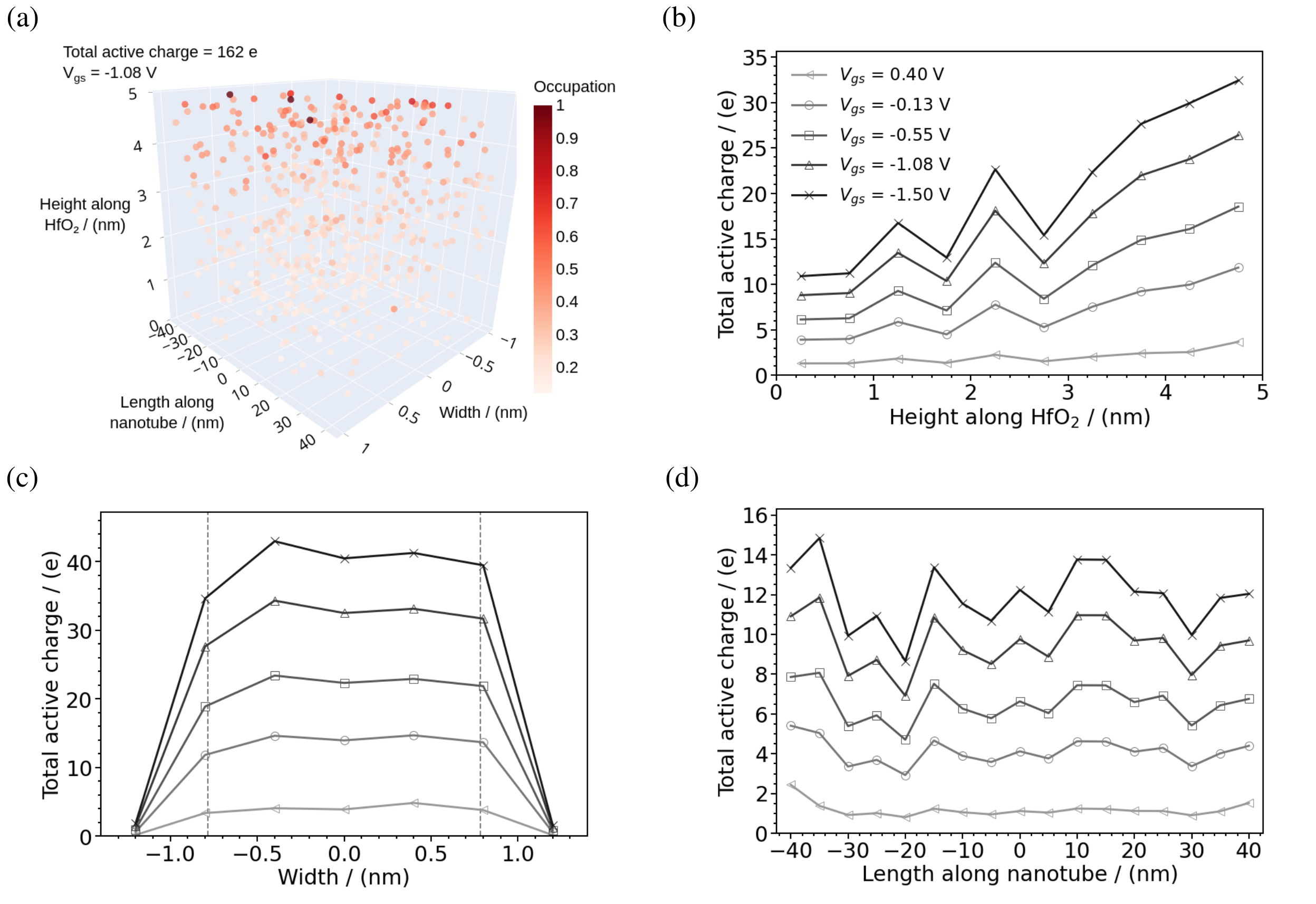}
       \caption
       {Characteristics of the 3D trap distribution at a trap density of 0.7 e/nm$^3$. (a) Trap occupation. (b)-d) Variation in total active charge across different gate-source voltages, as a function of: (b) height along HfO$_2$, (c) channel width, (d) length along the nanotube. Line plot legends for all panels are as shown in (b). In (c), the dashed vertical lines correspond to the width spanned by the nanotube.}
    \label{f:Fig6}
\end{figure*}
Figure~\ref{f:Fig5}a shows the calculated self-consistent band-bending across the CNT for a random 3D distribution with trap density of 0.7~e/nm$^3$ compared to the case without charge traps. The resulting $I-V_{\rm g}$ curves are displayed in Fig.~\ref{f:Fig5}b with charge traps leading to degraded performance (larger $SS$) similar to the experiment. 
We also performed simulations for 2D planar distributions with trap density of 1.6 e/nm$^2$, which showed similar large $SS$. The curves in Fig.~\ref{f:Fig5}b also show narrow standard deviations, obtained by carrying out 16 different simulations with different random seeds for the position of traps.
In all cases, the average size of randomly distributed clusters of traps with spacing $\Delta \leq 0.5$ nm is three, while the maximum cluster size is only eight.
This means that the aforementioned mechanism does not occur where bands get significantly distorted because of a large cluster size.
Notably, an additional band-bending due to traps is found (Fig.~\ref{f:Fig5}a) creating an additional barrier for hole injection and thus a lowered current at a given gate voltage.

The number of occupied traps (Fig.~\ref{f:Fig5}c) increases linearly as the gate voltage becomes more negative, as their energy level rises above the Fermi level. We also note that with a 2D planar distribution, we can obtain similar degraded performance with only one third the amount of total charge compared to the 3D distribution.
This is because only the traps closest to the CNT matter the most and many of the traps further away from the CNT are inconsequential, as we will discuss further below.

Figure~\ref{f:Fig5}d shows the effect of trap density on the $SS$ for both 3D and 2D distributions.
As the trap density increases, the $SS$ increases, eventually leading to good agreement with the experimental data. 
The degraded $SS$ can thus be explained by a linear gate-dependent charge trap occupancy, consistent with the expectation of Eq.~(\ref{eq:V_CNT}) with $N_{\rm it}$ proportional to $V_{\rm g}$. 
Figure~\ref{f:Fig5}d also shows the fraction of active charges at $V_{gs}=-1$~V, i.e. in the ON-state.
Note that the 3D density of 0.7 e/nm$^3$ and the 2D density of 1.6 e/nm$^2$ give a total number of 576 and 275 traps/CNT in the gate oxide, respectively.
However, given the active fraction of 0.28 and 0.21 at these densities, the amount of total active charges are 162 and 55, respectively.
For a uniform distribution this implies that charges need to be within ~0.4 nm of the CNT to have an appreciable effect.

The 2D trap charge value can be compared with those obtained with the continuum model; if we were to assume that the states are distributed over 1 eV, the density of trap states for the 2D distribution would be 2.6$\times 10^{13}$ e/cm$^2$/eV, not too dissimilar from experimental estimates based on capacitance measurements~\cite{Lin2023}. This can also be expressed as
\begin{equation}
    SS = SS_0 (1+\alpha\rho_{2D}),
    \label{eq:SSvsrho}
\end{equation}
with $\alpha = 0.96$ nm$^2$/e. Assuming a $\Delta E$ distribution of traps as before (i.e. $N_{\rm it}=\rho_{2D} / \Delta{E}$), Eq.~(\ref{eq:SS_2}) with $t_{\rm ox}=5$ nm and $\Delta E=1$ eV would predict $\alpha = 3.6$ nm$^2$/e, more then three times the simulated value. Narrower energy distributions would skew the agreement further. This highlights the need to utilize more specific simulations of discrete traps to quantitatively extract trap densities.

The mechanism by which the number of occupied traps increases with more negative gate voltage is illustrated in Fig.~\ref{f:Fig6} for the 3D trap distribution.
Figure~\ref{f:Fig6}a shows trap occupation within HfO$_2$ in the ON-state, with traps closest to the gate at the top almost fully active (occupation $\approx$ 1), while the occupation decreases away from the gate.
This is because the top layer of traps is directly affected by the gate potential, while the subsequent layers experience relatively less negative gate potential, leading to lower occupation.
This is seen in Fig.~\ref{f:Fig6}b, showing that the decrease in occupation, and therefore, the total active charge is nearly linear, with a steeper gradient of charge decrease within HfO$_2$ as the gate voltage becomes more negative. Figures~\ref{f:Fig6}c and \ref{f:Fig6}d show similar plots as a function of channel width and length along the nanotube, respectively. 
This shows that the most active traps lie within a width spanned by the nanotube, while the amplitude of trap charge variation along the nanotube increases as the gate voltage becomes more negative. Along the length of the CNT the trap occupancy is relatively uniform and increases linearly with gate voltage. 

\begin{figure}
    \centering
       \includegraphics[width=0.5\textwidth]{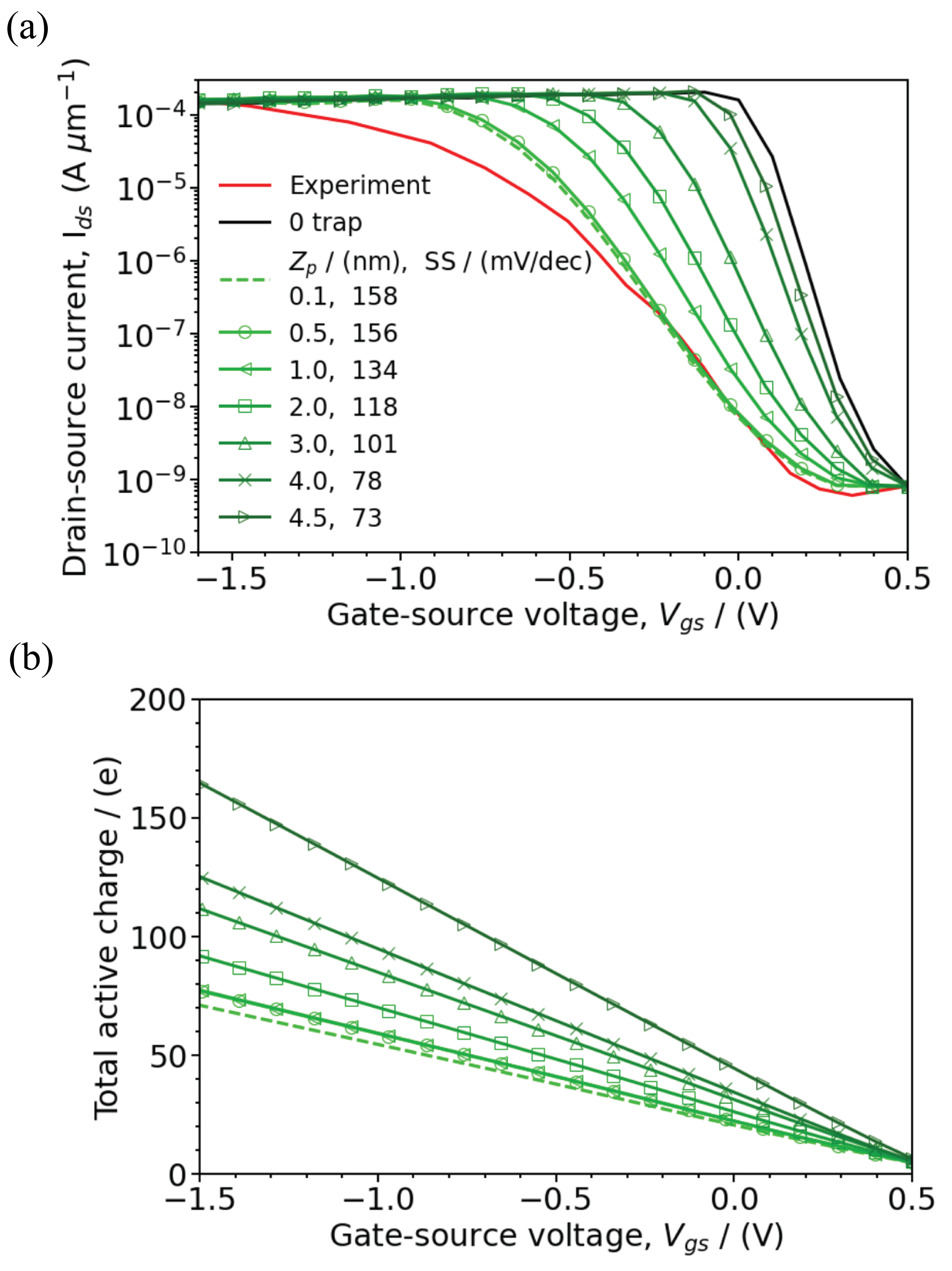}
       \caption
       {Effect of proximity of a 2D planar sheet of charge to the CNT array. (a) Variation in $I-V_{gs}$ characteristics along with corresponding SS.   (b) Variation in total active charge. For cases with different $Z_p$, the random seed was kept constant.}
       \label{f:Fig8}
\end{figure}

The result that traps closest to the gate have the highest occupancy may at first seem at odds with the results of Fig.~\ref{f:Fig5} where we showed that a 2D distribution of traps close to the CNT is the most important. This can be explained by the gate screening of the charge: a point charge at distance $d$ from the gate creates a dipole potential on the CNT proportional to $2d(h-d)^{-2}$ where $h$ is the gate oxide thickness, so with a 5 nm gate oxide thickness a charge 0.5 nm from the CNT is more than 7000 times more potent than one 0.5 nm from the gate. We verified this concept by simulating the impact of 2D distributions located at different distances from the CNT. Figure \ref{f:Fig8}a shows the impact on the transfer characteristics. As anticipated, traps 4.5 nm from the CNT make almost no change to the $SS$ while those close to the CNT have a strong effect. We also find a linear increase of the occupied charge with gate voltage (Fig.~\ref{f:Fig8}b), regardless of the proximity of the traps to the CNT. In addition, we find that degradation of the $SS$ occurs when around 50 traps per CNT are fully occupied.

\section{Conclusion}

In conclusion, progress in developing high-performance CNT transistors has brought the issue of imperfections to the forefront. In particular, charge traps have been identified as a potential source for the degraded subthreshold swing observed experimentally. The availability of new powerful simulation tools now allows for the detailed simulations of nanomaterials-based transistors with full 3D geometry with the ability to introduce variable occupancy discrete charge traps over the full 3D oxide. Using this approach, we show that discrete charge traps can explain recent experimental data on top-gate devices, with the charge traps closest to the CNTs being the most important. Future efforts to improve the performance of experimental systems will require the removal of a few tens of traps per CNT. More generally, the results presented here should apply to a broad range of nanomaterials-based electronic and optoelectronic devices, as well as to other sources of charge traps such as residual surfactant.

\begin{acknowledgements}
Work by S. S.~Sawant, A.~Nonaka, and Z.~Yao was supported by the U.S.~Department of Energy, Office of Science, Office of Applied Scientific Computing Research, the Microelectronics Co-Design Research Program (Co-Design and Integration of Nano-sensors on CMOS), and the Microelectronics Science Research Center (Nanoscale Hybrids: A New Paradigm for Energy-efficient Optoelectronics) under Contract DE-AC02-05-CH11231. T.~Lara was supported by the U.S.~Department of Energy, Office of Science Undergraduate Laboratory Internships (SULI) Program. 
This manuscript has been authored by employees of Lawrence Berkeley National Laboratory under Contract No. DE-AC02-05CH11231 with the U.S.~Department of Energy. The U.S.~Government retains, and the publisher, by accepting the article for publication, acknowledges, that the U.S. Government retains a non-exclusive, paid-up, irrevocable, world-wide license to publish or reproduce the published form of this manuscript, or allow others to do so, for U.S.~Government purposes.
Work by F.~L\'eonard was supported by the U.S.~Department of Energy, Office of Science, the Microelectronics Co-Design Research Program, under Contract DE-NA-0003525 (Co-Design and Integration of Nano-sensors on CMOS). Sandia National Laboratories is a multimission laboratory managed and operated by National Technology and Engineering Solutions of Sandia, LLC., a wholly owned subsidiary of Honeywell International, Inc., for the U.S. Department of Energy’s National Nuclear Security Administration
under Contract DE-NA-0003525. This written work is authored by an employee of NTESS. The
employee, not NTESS, owns the right, title and interest in and to the written work and is
responsible for its contents. Any subjective views or opinions that might be expressed in the written
work do not necessarily represent the views of the U.S. Government. The publisher acknowledges
that the U.S. Government retains a non-exclusive, paid-up, irrevocable, world-wide license to
publish or reproduce the published form of this written work or allow others to do so, for U.S.
Government purposes. The DOE will provide public access to results of federally sponsored
research in accordance with the DOE Public Access Plan. 
Department of Energy's National Nuclear Security Administration under contract DE-NA-0003525.
This research used resources of the National Energy Research Scientific Computing Center (NERSC), a Department of Energy Office of Science User Facility using NERSC award ASCR-ERCAP0026882.
\end{acknowledgements}

\section*{Data availability}

Data to support the conclusions of this work is available from the authors upon reasonable request.

\section*{Code availability}~

The code used in this work, ELEQTRONeX, is open-source and available in the GitHub repository maintained by the ELEQTRONeX team: \href{https://github.com/AMReX-Microelectronics/ELEQTRONeX}{https://github.com/AMReX-Microelectronics/ELEQTRONeX}. Documentation for ELEQTRONeX can be found at \href{https://amrex-microelectronics.github.io/}{https://amrex-microelectronics.github.io/}. Use of this software is under a modified BSD license – the license agreement is included in the ELEQTRONeX home directory as license.txt

\bibliography{references}

\end{document}